\documentclass[prb,aps,10pt,twocolumn,showkeys,showpacs,amsmath,amsfonts,floatfix,superscriptaddress,nofootinbib]{revtex4-2}
\usepackage{amssymb,amsmath,amstext}
\usepackage{graphicx}
\usepackage{epstopdf}
\usepackage{xcolor}
\usepackage{bm}
\usepackage[utf8]{inputenc}
\usepackage{ulem}
\usepackage{latexsym}
\usepackage[colorlinks=true,citecolor=blue,linkcolor=magenta]{hyperref}
\usepackage{xr-hyper}
\def\be{\begin{equation}}
\def\ee{\end{equation}}
\def\bea{\begin{eqnarray}}
\def\eea{\end{eqnarray}}
\def\bi{\begin{itemize}}
\def\ei{\end{itemize}}

\newcommand{\bra}[1]{\mbox{$\langle #1 |$}}
\newcommand{\ket}[1]{\mbox{$| #1 \rangle$}}
\newcommand{\braket}[2]{\mbox{$\langle #1  | #2 \rangle$}}

\newcommand{\affildoc}{
             Jagiellonian University,
             Doctoral School of Exact and Natural Sciences,
             ul. \L{}ojasiewicza 11, 30-348 Krak\'ow, Poland
             }

\begin{document}



\title{Finite temperature dopant-induced spin reorganization explored via \\ tensor networks in the two-dimensional $t$-$J$ model}

\author{Yintai Zhang}
\affiliation{\affildoc}
\affiliation{Jagiellonian University,
            Faculty of Physics, Astronomy and Applied Computer Science,
            Institute of Theoretical Physics,
            ul. \L{}ojasiewicza 11, 30-348 Krak\'ow, Poland}

\author{Aritra Sinha}
\affiliation{Max Planck Institute for the Physics of Complex Systems,
            N\"{o}thnitzer Stra\ss e 38, Dresden 01187, Germany}

\author{Marek M. Rams}
\affiliation{Jagiellonian University,
            Faculty of Physics, Astronomy and Applied Computer Science,
            Institute of Theoretical Physics,
            ul. \L{}ojasiewicza 11, 30-348 Krak\'ow, Poland}
\affiliation{Jagiellonian University,
            Mark Kac Center for Complex Systems Research,
            ul. \L{}ojasiewicza 11, 30-348 Krak\'ow, Poland}

\author{Jacek Dziarmaga}
\affiliation{Jagiellonian University,
            Faculty of Physics, Astronomy and Applied Computer Science,
            Institute of Theoretical Physics,
            ul. \L{}ojasiewicza 11, 30-348 Krak\'ow, Poland}
\affiliation{Jagiellonian University,
            Mark Kac Center for Complex Systems Research,
            ul. \L{}ojasiewicza 11, 30-348 Krak\'ow, Poland}

\date{\today}

\begin{abstract}
We study the two-dimensional $t$--$J$ model at finite temperature directly in the thermodynamic limit using purification represented by an infinite projected entangled-pair state (iPEPS). We reach temperatures down to $T/t=0.1$ and hole concentrations up to $1-n\simeq0.25$, and provide benchmark thermodynamic-limit results for the specific heat, uniform susceptibility, and charge compressibility. We identify a susceptibility maximum $T^\ast$ that tracks the buildup of short-range antiferromagnetism and a shallow compressibility enhancement upon cooling in the same doping window. To expose the underlying microscopic mechanism, we introduce dopant-conditioned multi-point correlators that quantify how holes reorganize nearby exchange: single holes weaken adjacent antiferromagnetic bonds, while nearest-neighbor hole pairs produce a cooperative response that reinforces antiferromagnetism on the parallel plaquette edge. Over the same parameter window, $d$-wave pairing correlations remain short-ranged. These results provide experiment-compatible thermodynamic-limit benchmarks and establish dopant-conditioned correlators as incisive probes of finite-temperature spin-texture reorganization in doped Mott insulators.
\end{abstract}
\maketitle

\section{INTRODUCTION}
\label{sec:intro}
Strongly correlated electron systems on a two-dimensional (2D) lattice exhibit a spectacular variety of collective behaviors, all rooted in three simple ingredients: the tendency of electrons to hop between sites (kinetic energy), their magnetic superexchange interactions, and thermal fluctuations. In materials ranging from the cuprates and nickelates to layered ruthenates and twisted-bilayer graphene~\cite{keimer15,li2019nickelate,mackenzie2003srt2ruo4,cao2018sc,cao2018ci}, doping a Mott-insulating antiferromagnet gives rise to intertwined spin and charge textures often organizing into stripes~\cite{fradkin2015, keimer15} and to unconventional superconductivity with $d$-wave or other exotic pairing symmetries~\cite{Lee2006}. As temperature rises, these same systems enter the enigmatic pseudogap regime, characterized by a partial suppression of the electronic density of states~\cite{timusk1999pseudogap}, and on further doping the strange-metal regime, where transport violates conventional Fermi-liquid rules~\cite{phillips2022}.

To capture and understand this rich phenomenology in its simplest form, theorists have long turned to two paradigmatic lattice Hamiltonians: the single-band Hubbard model~\cite{Hubbard63}, which supports hopping to local neighbors against on-site Coulomb repulsion, and its strong-coupling limit, the $t$-$J$ model~\cite{tJ_Oles}, which eliminates double occupancy and explicitly encodes spin-exchange interactions. Despite their simplicity, these models have been immensely successful at reproducing key features of high-temperature superconductors including stripes, polarons, superconductivity, and non-Fermi-liquid behavior.

At zero temperature, extensive numerical studies of the Hubbard and $t$-$J$ models using density-matrix renormalization group~\cite{white98_tJ, qin20}, infinite projected entangled pair states~\cite{corboz14_tJ, zheng17}, and large-scale multi-method benchmarks~\cite{zheng17} combining DMRG, auxiliary-field quantum Monte Carlo, density matrix embedding theory (DMET) and other variational approaches have established that these minimal Hamiltonians with only nearest-neighbor coupling host stripe-ordered phase in the under-doped regime. Recent DMRG studies conducted on wide cylinder geometries have found robust $d$-wave superconductivity, possibly coexisting with pair density wave~\cite{haldane2025}. Other advances include fermionic projected entangled pair states finding evidence for stable diagonal stripes~\cite{dong2020stable} and pair-density-waves competition in the square-lattice $t$-$J$ model~\cite{Zheng2024}, and neural quantum state simulations of the $t$-$J$ model at finite doping~\cite{Lange2024}. In parallel, time-dependent studies show that non-equilibrium quenches can transiently expose stripe growth and phase-separation tendencies~\cite{Luke2025}. The stripe order often compete or co-exist with $d$-wave pairing correlations when parameters such as the next-nearest-neighbor hopping are tuned~\cite{Boris2019, jiang2021, xu2024,jiang2023sixleg, Lu2025, Qu2024}.

Extending these studies to finite temperature poses severe challenges for classical computation. Exact diagonalization and finite-temperature Lanczos methods (FTLM) have captured the onset temperature of pseudogap through maxima in uniform magnetic susceptibility on small clusters~\cite{jaklivc2000}, while determinantal and diagrammatic quantum Monte Carlo have mapped out the evolution of spin and charge correlations in the pseudogap and strange-metal regime~\cite{huang2023, simkovic2024}. Tangent-space tensor renormalization (tanTRG) enables controlled finite-temperature simulations on wide Hubbard cylinders and large square lattices, revealing low-temperature doped-regime behavior and pseudogap signatures in Matsubara Green's functions~\cite{Li2023}. In thin cylinder geometries, minimally entangled typical thermal states (METTS) simulations unraveled a finite-temperature crossover to the stripe order~\cite{wietek2021}, marked by anomalies in thermodynamics and recently the presence of hole clusters surrounded by antiferromagnetic domains at finite temperature~\cite{sinha2024forestalled}. New finite-$T$ probes complement this picture: one-hole spectroscopy in the $t$-$J$ model resolves magnetic-polaron features and their thermal evolution~\cite{Guthardt2025}, while numerical linked-cluster expansion approaches report finite-$T$ kinetic-ferromagnetism regimes~\cite{newby2025}, and strange metal transport through FTLM~\cite{Fratini2025}. Yet all these methods either encounter prohibitive sign-problems at strong coupling and low temperature or suffer from finite-size limitations that obscure the true thermodynamic behavior.

Ultracold-atom quantum simulators have emerged as a powerful experimental platform for realizing the Fermi-Hubbard Hamiltonian with tunable interactions, doping, and temperature. Quantum-gas microscopes~\cite{Bakr2009} now provide single-site, spin-resolved snapshots of fermions in optical lattices, directly imaging antiferromagnetic domains, hidden string order, and polaronic textures at finite temperature~\cite{Parsons2016, chiu19, koepsell2021microscopic}. Recent work combined such site-resolved measurements with multi-point correlator analysis up to fifth order, uncovering a universal, doping-dependent temperature scale that tracks the onset of the pseudogap through emergent scaling of spin–charge correlations~\cite{Chalopin2024}. Most recently, a cryogenic Hubbard simulator was realized reaching unprecedentedly low temperatures $T/t\lesssim0.1$ extending low-temperature control into the doped regime~\cite{xu2025neutral}. Building on that, Ref.~\cite{Kendrick2025Pseudogap} observed a crossover to a pseudogapped metal in a Fermi–Hubbard quantum simulator: the compressibility develops a doping-tuned maximum and traces a line of thermodynamic anomalies across interaction strength, while lattice-modulation spectroscopy reveals a momentum-selective depletion of low-energy spectral weight.

Inspired by these advances, we use purification~\cite{verstraete2004matrix,czarnik12,CzarnikDziarmagaCorboz} with infinite projected entangled pair states (iPEPS)~\cite{jordan2008,corboz2010,czarnik15b,ntu} ansatz to study the two dimensional square lattice $t$-$J$ model at finite temperature directly in the thermodynamic limit, with significant technical improvements over the technique applied to a previous study on the Fermi-Hubbard model~\cite{Hubbard_Sinha}. In this paper we make three contributions. 
First, we provide thermodynamic-limit finite-$T$ benchmarks for the square-lattice $t$--$J$ model down to $T/t=0.1$ across dopings up to $1-n\simeq0.25$. Second, we extract and track the susceptibility maximum $T^\ast$ and a shallow compressibility enhancement upon cooling, which together delineate the regime of strongest short-range spin--charge interplay in our data.
Third, we introduce dopant-conditioned three- and four-point correlators that directly visualize how single holes and nearest-neighbor hole pairs reorganize local exchange, and we relate these microscopic responses to the thermodynamic anomalies.

\section{Model}
\label{sec:model}
The two-dimensional $t$-$J$ model, originally derived from the Hubbard model in the large-$U$ limit via a Schrieffer-Wolff transformation\cite{tJ_Oles}, has long served as a cornerstone for understanding doped Mott insulators. In particular, it captures essential ingredients believed to be relevant to high-$T_c$ cuprate superconductors, including strong antiferromagnetic (AFM) correlations at half-filling and the interplay of spin and charge degrees of freedom upon doping. We consider the standard $t$-$J$ Hamiltonian on a square lattice,
\begin{align}
H \;=\;
& -t \sum_{\langle i,j \rangle,\sigma}
\Bigl(
\tilde c_{i\sigma}^\dagger \,\tilde c_{j\sigma} + \tilde c_{j\sigma}^\dagger \,\tilde c_{i\sigma}
\Bigr) \nonumber\\[2mm]
&+\; J \sum_{\langle i,j\rangle}
\Bigl(
\mathbf{S}_i \cdot \mathbf{S}_j - \tfrac{n_i n_j}{4}
\Bigr) -\; \mu \sum_{i} n_i,
\label{eq:tJ-Ham}
\end{align}
where the operator $\tilde c_{i\sigma}$ projects out double occupancy, $\mu$ is the chemical potential, and $n_i$ is the electron number operator. In all numerical simulations, we control doping by tuning $\mu$ and measure the filling $n$. Unless stated otherwise, all quantities are plotted vs. temperature $T$ in units of the hopping $t=1$. The filling $n$ runs from $1$ (half-filled) down to $\approx 0.75$. In the following we set $J=0.5$. 
\begin{figure}[t!]
\centering
    \includegraphics[page=1]{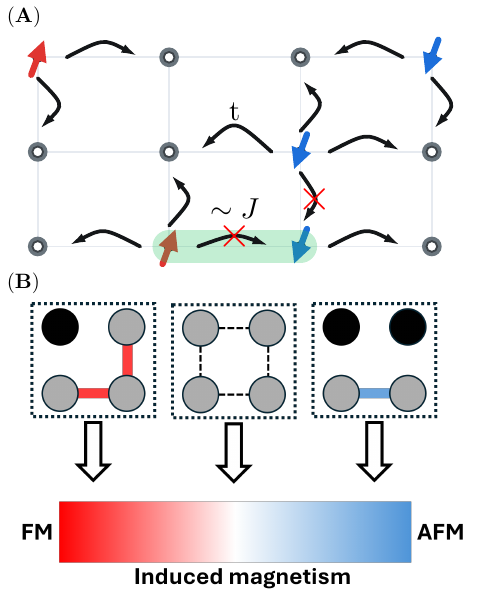}
    \caption{\textbf{Schematics of \textit{t-J} model dynamics and dopant-induced spin correlations -- }(A) Schematic of the two-dimensional $t$-$J$ model on a square lattice with the nearest neighbor hopping $t$ and exchange interaction $J$. In a representative classical picture, up (red) and down (blue) spins can hop around to sites with holes (gray circles) while double occupation of single lattice site is forbidden. (B) Hole-conditioned spin-spin correlation. Nearest neighbor (NN) bonds are antiferromagnetic (AFM; NN $\langle S^{z}_{i}S^{z}_{j}\rangle <0$) and conditioned correlation measure where we remove the background AFM correlation is trvially $0$ as we see in the middle dotted box. The left dotted box shows that conditioned on a single hole at the black site, the NN AFM bonds nearby are weakened and become more ferromagnetic compared to the background (represented by red solid lines). With two adjacent holes, AFM is reinforced on the plaquette edge parallel to the pair, i.e., the corresponding NN bonds become more antiferromagnetic compared to the background (represented by the blue solid line) as shown in the right dotted box.}
\label{fig:Fig1}
\end{figure}

Fig.~\ref{fig:Fig1} illustrates the dynamics of the $t$-$J$ model and the principal physics in the article. Panel A is a classical sketch of the doped $t$-$J$ model lattice: spins (up: red; down: blue) occupy some lattice sites, other lattice sites have holes denoted by empty gray circles. Here, allowed moves are spin hops into a neighboring hole occupied site, and any move that would create double occupancy is forbidden (marked with a red $\times$); bonds connecting two occupied sites with opposite spins are highlighted with a green halo to indicate that antiparallel occupied bonds have superexchange contribution $\sim J$, while bonds touching a hole or like spins carry no exchange contribution. Panel B shows how dopants reorganize the local magnetism of an antiferromagnetic background. The three scenes correspond to the unconditioned AFM background (center), one fixed hole (left), and two neighboring fixed holes (right). A single hole (a magnetic polaron) disturbs nearby AFM links and produces a short-range ferromagnetic skew around it, whereas two adjacent holes moving together break fewer AFM bonds than two isolated holes and therefore cooperatively stabilize an antiferromagnetic segment along the plaquette edge parallel to the pair.

The method of choice here is purification within an infinite projected entangled-pair state ansatz, which has been previously successfully applied to a variety of problems~\cite{CzarnikDziarmagaCorboz, kshetrimayum19, czarnikKH, czarnikSS, jimenez2021quantum, poilblanc_Heisenberg, plateau_melting}. The purification is evolved in imaginary time from $T=\infty$ to a given temperature. The ansatz uses a two-site checkerboard with $U(1)\times U(1)$ symmetry. We optimize imaginary-time gates using a neighborhood tensor update~\cite{ntu,mbl_ntu,Banuls2014} with an environment assisted initial truncation~\cite{Hubbard_Sinha}. We evaluate observables with a multiplet-aware extension of the corner transfer matrix renormalization group~\cite{orus2009, nishino1996,corboz14_tJ}. Working directly in the thermodynamic limit avoids finite-size bias. The correlations are then computed using the zipper algorithm~\cite {METTS_PEPS}. Please check Appendix \ref{app:numerical_details} for methodological and numerical details. All simulations were performed with the YASTN library~\cite{10.21468/SciPostPhysCodeb.52, 10.21468/SciPostPhysCodeb.52-r1.2}.

\section{Thermodynamics}\label{sec:thermo}
\begin{figure}[htb!]
    \includegraphics[page=2]{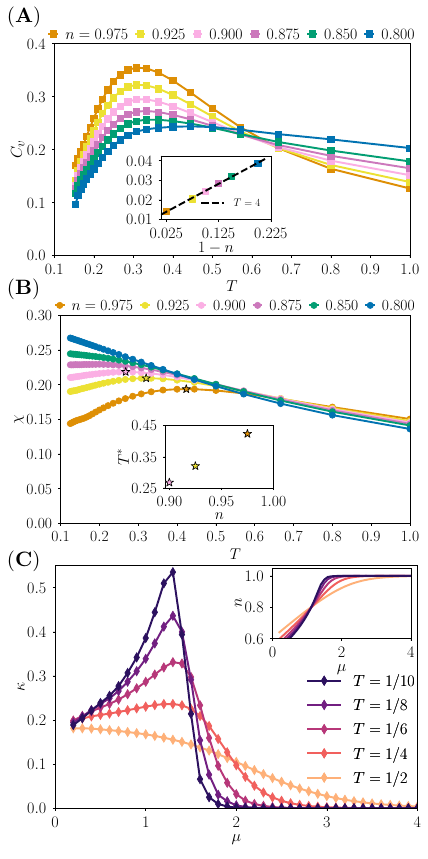}
    \caption{\textbf{Thermodynamic observables -- }(A) shows specific heat $C_v(T)$. Inset: $C_v$ linearly depends on the doping $1-n$ in the high-temperature regime as predicted in Ref.~\cite{rigol2007}. (B) shows uniform spin susceptibility $\chi(T)$, obtained from Eq.~(\ref{eq:chi_est}). (C) shows charge compressibility $\kappa = \partial n/\partial \mu$ versus $\mu$ at several fixed temperatures. The inset shows $n(\mu)$ vs $\mu$.
}
\label{fig:Thermodynamics}
\end{figure}

\vspace*{0.5em}
\noindent Fig.~\ref{fig:Thermodynamics} summarizes bulk thermodynamics of the square lattice $t$-$J$ model in the thermodynamic limit. We focus especially on the low-to-intermediate temperature range $0.1 \lesssim T \lesssim 0.5$, where short-range AFM order and emergent pseudogap features become prominent~\cite{prelovvsek2000pseudo}. Fig.~\ref{fig:Thermodynamics}(A) shows the specific heat $C_{v}(T)=\partial_{T}{\langle H\rangle}$, which exhibits the expected spin-entropy peak associated with the formation of short-range antiferromagnetism. At (or very near) half filling ($n=1$), a broad maximum appears at $T\sim J$, reflecting the release of entropy when nearest-neighbor singlets proliferate and spin correlations grow. Upon increasing doping $1-n$, the peak diminishes in height and shifts to higher $T$, consistent with mobile holes breaking AFM bonds, reducing the spin stiffness, and shortening the antiferromagnetic length. In the high temperature window, $T \geq 1$, the inset reveals the anticipated linear in $1-n$ behavior of $C_{v}$ from high $T$ expansion: energy fluctuations are dominated by uncorrelated spins and holes. Please refer to Appendix \ref{app:chemical_potential} to see how doping depends on chemical potential and temperature $T$.

Fig.~\ref{fig:Thermodynamics}(B) displays the uniform spin susceptibility $\chi(T)$. In principle,
\begin{equation}
\chi = \frac{1}{NT}\sum_{i,j} \langle S_i^z\,S_j^z\rangle=
\frac{\langle (S^z_{\rm tot})^2\rangle}{N\,T},
\label{eq:chi_0}
\end{equation}
with $S^z_{\text{tot}}=\sum_i S^z_i$. However, in an infinite PEPS the direct evaluation of Eq.~(\ref{eq:chi_0}) is ill-conditioned at low $T$: the slow decay of $\langle S^z_i S^z_j\rangle$ with $|i{-}j|$ makes the sum sensitive to tiny long-distance errors. We therefore obtain $\chi$ from its definition. Adding a small uniform $S^{z}$ field,
\begin{equation}
H \to H - h\sum_i S^z_i,\qquad m(h)=\frac{1}{N}\sum_i \langle S^z_i\rangle,
\end{equation}
we extract
\begin{equation}
\chi = \left.\frac{d m}{d h}\right|_{h\to0}\;\approx\;\frac{m(h)}{h}.
\label{eq:chi_est}
\end{equation}

\begin{figure*}[htb!]
\centering
    \includegraphics[page=3]{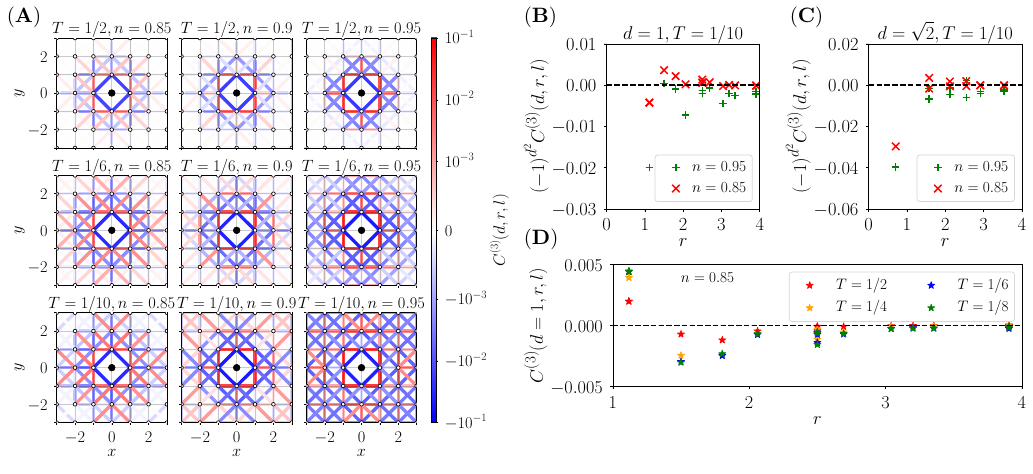}
     \caption{\textbf{3P correlator -- } (A) shows 3P correlators of different temperatures and fillings in a $7\times7$ window. The black dot sitting at the center indicates the position of the hole projector $h$. $S^z$ operators are applied on the nearest neighboring sites (with a distance of $d=1$) or the next nearest neighboring sites (with a distance of $d=\sqrt{2}$). The intensities of the colors of different segments represent the magnitudes of different 3P correlators. The segment is red (blue) if the corresponding correlator is positive (negative). The correlators whose absolute values are smaller than $10^{-3}$ are shown on a linear scale, while  others are on a logarithmic scale. The range of the 3P correlators increases when the temperature is lowered for a fixed filling. The signs of some long-range correlators alter when changing the filling or the temperature.
     (B) and (C)
     show how the correlators depend on $r$ at $T=1/10$, which represents the distance between the hole and the bond linking two $S^z$ operators of the 3P correlator, for $d=1$ and $d=\sqrt{2}$ respectively.
     (D)
     shows the $r$ dependence of $d=1$ at various temperatures for filling $n=0.85$.
}\label{fig:3P_large}
\end{figure*}

\noindent In this work, we set $h=0.1$ and $h=0.01$. The results of these fields match with each other quantitatively, proving that we are in the weak-field regime (see Appendix~\ref{app:different_field}). For weak-doping regime (e.g., $n \sim 0.95$), $\chi(T)$ develops a clear maximum at a temperature $T^{*}$ (see the inset of Fig.~\ref{fig:Thermodynamics}(B)), which tracks the build-up of strong but short-ranged AFM correlations: cooling below $T^{*}$ suppresses spin fluctuations as local singlets form; heating above $T^{*}$ melts the AFM background into a Curie-like regime. As doping increases, the maximum weakens and shifts to lower $T$. We thus identify $T^{*}$ as a pseudogap crossover scale set by short-range~\cite{johnston1989magnetic, wietek2021}. Its evolution mirrors the suppression of the $C_{v}$ peak and anchors the microscopic reorganization discussed later. In Appendix \ref{app:ipeps_vs_metts}, we compare our iPEPS results for specific heat and magnetic susceptibility against METTS using matrix product states applied to cylinders and have appreciable agreement.

Fig.~\ref{fig:Thermodynamics}(C) shows the isothermal compressibility $\kappa(\mu,T)=(\partial n/\partial\mu)_T$ together with filling $n(\mu)$ (inset). At high temperature, the response is nearly featureless, as expected for a classical mixture of spins and holes with weak correlations. Upon cooling to about $T\sim 0.3$ range, $\kappa$ develops a small peak as a function of $\mu$ (or equivalently $n$). This emerges in the same doping window where $\chi(T)$ exhibits a pronounced maximum. Two competing trends shape $\kappa$ here. Close to half filling, Mott constraints suppress charge fluctuations and reduce $\kappa$; moving away from half filling, the growth of short-range AFM domains lowers the energy cost of rearranging charge locally (e.g., by accommodating holes so as not to destroy the background AFM), mildly enhancing $\kappa$. The net result is a small but robust peak which grows as a function of decreasing temperature, consistent with a crossover --- not a phase transition --- in two dimensions at finite $T$. This intermediate-$T$ peak of $\kappa$ versus doping mirrors the cluster-DMFT Widom-line phenomenology of the doped Mott insulator, where $\kappa$ and related thermodynamic derivatives reach maxima above a pseudogap–metal critical endpoint and thus track the pseudogap boundary~\cite{khatami2010, sordi2011mott, sordi2012pseudogap}. This phenomenon has also been tied to charge clustering~\cite{sinha2024forestalled} by one of the authors, and our METTS snapshots on a cylindrical lattice in Appendix~\ref{app:clustering} show the same growth of fluctuating hole clusters without macroscopic phase separation. There, we detail the analysis: we flag hole-rich sites via an adaptive threshold, group them into nearest-neighbor connected components on a cylinder of length $24$ and width $4$ with periodic boundary conditions, and compile density-weighted cluster-size histograms resolved by total hole mass. As the temperature is lowered, the probability for having clusters of a certain size shifts toward larger but still finite clusters. Our compressibility values also agree qualitatively with the finite-temperature Lanczos benchmarks of Ref.~\cite{jaklivc2000}. A recent Fermi–Hubbard quantum simulator study similarly mapped a pseudogap phase diagram via a doping-tuned compressibility maximum and additionally momentum-selective depletion of low-energy spectral weight, revealing a line of thermodynamic anomalies and suggesting a link to charge order~\cite{Kendrick2025Pseudogap}.

\section{Higher Order Correlators}
\label{high_order_correlators}
\vspace*{0.5em}

\begin{figure}
    \centering
    \includegraphics[width=8.5cm]{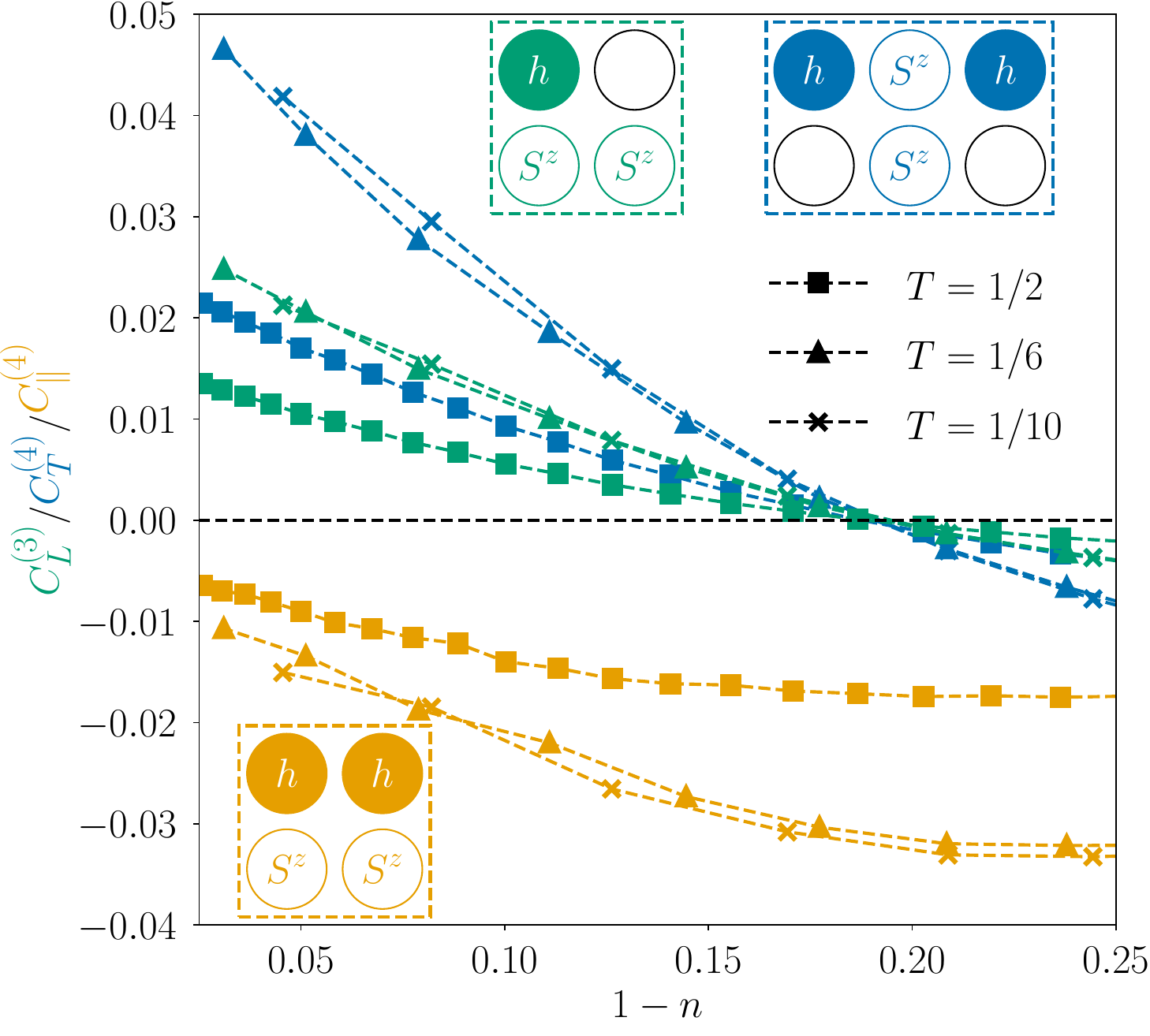}
    \caption{\textbf{3P correlator versus 4P correlator -- } Green, blue, yellow insets show the configuration of $C^{(3)}_L$, $C^{(4)}_T$, and $C^{(4)}_\parallel$ respectively. Different markers represent different temperatures. Different correlators are plotted in accordance with the color of the corresponding insets.}
    \label{fig:3Pvs4P}
\end{figure}

To quantify how dopants reorganize short range magnetism beyond two-point functions, we analyze conditional spin-charge correlators inspired by studies done in Refs.~\cite{ Koepsell2019, blomquist2020unbiased, blomquist2021evidence, koepsell2021microscopic, Chalopin2024}. Let $h_{i} \equiv 1- n_{i}$ be the hole projector and let $B_{ab} \equiv S_{a}^{z} S_{b}^{z}$ denotes the $zz$-component of the spin-spin operator on the bond $(a,b)$. For any disjoint site sets $A,B$ and operators $\hat{\mathcal O}_A$, $\hat{\mathcal P}_B$ (with $\hat{\mathcal P}_B$ a projector), we define the normalized conditional correlator
\begin{equation}
\mathcal{C}[\hat{\mathcal O}_A \mid \hat{\mathcal P}_B]
\;\equiv\;
\frac{\langle \hat{\mathcal O}_A \hat{\mathcal P}_B\rangle}{\langle \hat{\mathcal P}_B\rangle}
\;-\;
\langle \hat{\mathcal O}_A\rangle,
\label{eq:cond_def}
\end{equation}
i.e. the change in the expectation of $\hat{\mathcal O}_A$ when conditioning on $\hat{\mathcal P}_B$ relative to the unconditional background. In practice we use (i) three-point (3P) correlators
\begin{equation}
C^{(3)}_{ab\,|\,i}\equiv \mathcal{C}[B_{ab}\mid h_i]
=\frac{\langle B_{ab}\,h_i\rangle}{\langle h_i\rangle}-\langle B_{ab}\rangle,
\label{eq:C3_def}
\end{equation}
and (ii) four-point (4P) correlators with two holes,
\begin{equation}
C^{(4)}_{ab\,|\,ij}\equiv \mathcal{C}[B_{ab}\mid h_i h_j]
=\frac{\langle B_{ab}\,h_i h_j\rangle}{\langle h_i h_j\rangle}-\langle B_{ab}\rangle.
\label{eq:C4_def}
\end{equation}
We parameterize a 3P observable by the bond length $d=\|\mathbf{r}_{a}{-}\mathbf{r}_{b}\|$, bond-hole separation $r=\big\|(\mathbf{r}_{a}+\mathbf{r}_{b})/2-\mathbf{r_i}\big\|$, and the shorter of the two separations between a site applied with a spin operator and the hole $l$. The shortest correlator we monitor is the ``L-shape'' $C^{(3)}_{L}\equiv C^{(3)}(d{=}1,\,r{=}\tfrac{\sqrt5}{2},\,l{=}1)$ (see the green inset of Fig.~\ref{fig:3Pvs4P}). For 4P we focus on two inequivalent arrangements: (i) a unit-square with parallel bonds, where the two holes occupy nearest neighbors and the probed spin bond is the parallel edge on the adjacent side, denoted $C^{(4)}_{\parallel}$ (see the orange inset of Fig.~\ref{fig:3Pvs4P}); and (ii) a ``$T$'' geometry where the holes are separated by two lattice spacings and the probed nearest-neighbor bond is centered between them along the perpendicular direction, denoted $C^{(4)}_{T}$ (see the blue inset of Fig.~\ref{fig:3Pvs4P}). Note that 4P correlators are more sensitive to numerical noise because the numerical accuracy is reduced by $\langle h_i h_j\rangle\leq(1-n)^2$.
The subtraction of two close numbers in \eqref{eq:cond_def} are sensitive to both numerical and interpolation errors. To avoid the latter, we first solve $n(T,\mu_0)=n_0$ with respect to $\mu_0$ for a given filling $n_0$. Then the whole ${\cal C}(T,\mu)$ is interpolated to $\mu_0$ rather than the three expectation values on the right hand side of \eqref{eq:cond_def} individually.

In Fig.~\ref{fig:3P_large}, we evaluate the spin-spin-hole 3P correlators for bonds at distance $d=1$ or $\sqrt{2}$. In a N\'eel background the unconditional correlator $\langle B_{ab}\rangle$ is negative for nearest neighbours ($d{=}1$) and positive for 
diagonals ($d{=}\sqrt{2}$). Hence a red (positive) $C^{(3)}$ on $d{=}1$ means the AFM bond is weakened (made less negative) by the hole, whereas a blue 
(negative) $C^{(3)}$ on $d{=}\sqrt{2}$ means the originally positive diagonal correlation is suppressed, both indicating a local FM effect of the 
hole. At high to intermediate temperatures ($T\geq 1/4$), the correlation maps are short-ranged across $n=0.85$, $0.9$ and $0.95$, but grow upon cooling and with increasing filling, as shown in 
Fig.~\ref{fig:3P_large}(A). At lower temperature $T=1/10$, the correlator extends over the whole $7\times7$ window for all fillings with a much slower decay. Figs.~\ref{fig:3P_large}(B), (C), and (D) show these general tendencies in more detail. 
Panels (B) and (C) quantify the spatial dependence of the 3P response at fixed $T/t=0.1$ as a function of the bond--hole separation $r$, for nearest-neighbor bonds ($d=1$) and diagonal bonds ($d=\sqrt{2}$), respectively. Upon increasing filling (reducing 
doping), the magnitude and the spatial range of the conditioned response increase, consistent with the growth of short-range AFM correlations in the background. Panel (D) shows the temperature evolution at fixed filling $n=0.85$: cooling extends the range of the 3P response and increases its magnitude, while the high-$T$ curves remain short-ranged. These panels make explicit that the dopant-conditioned spin response is primarily a short- to intermediate-range reorganization that strengthens markedly as AFM correlations build up on cooling.

In Fig.~\ref{fig:3Pvs4P} we contrast the single-dopant L-geometry $C^{(3)}_{L}$ with two different two-dopant conditionings: the well separated $T$ configuration $C^{(4)}_{T}$, and the nearest-neighbor plaquette configuration $C^{(4)}_{\parallel}$ as a function of the doping $1-n$. For dopings $0.00 \lesssim 1-n \lesssim 0.18$, both $C^{(3)}_{L}$ and $C^{(4)}_{T}$ are positive, then cross zero near $n = 0.8$ and become negative upon further doping. The change in sign can be attributed to weakening of AFM correlations due to increased holes. Moreover, the near-additivity $C^{(4)}_{T} \simeq 2 C^{(3)}_{L}$ holds on both sides of the crossing within the plotted scatter, indicating weak interaction between the well-separated dopants. Interpreting the sign relative to the background bond value, the positive regime signals a local ferromagnetic skewing around a dopant (AFM bonds made less negative), whereas beyond filling $n \sim 0.8$, the softened AFM background (reduced spin stiffness) leads the conditioned response to reinforce the residual AFM correlations (AFM bonds made more negative). By contrast, $C^{(4)}_{\parallel}$ is
negative across the entire range and grows in magnitude away from half filling (increased doping $1-n$) and on cooling, consistent with the following mechanism: two adjacent holes eliminate seven, rather than eight, $J$–links, lowering the exchange cost and cooperatively enforcing AFM on the parallel edge of the plaquette. This response, at finite $T$, provides a microscopic route toward incipient charge segregation marked by enhanced compressibility (as shown in Fig.~\ref{fig:Thermodynamics}(C)); upon further cooling, it can act as a precursor to stripe formation where holes line up to minimize exchange frustration while retaining kinetic energy. Finally, the doping where $C^{(3)}_{L}$ and $C^{(4)}_{T}$ change sign coincides with the regime where the uniform susceptibility peak fades in thermodynamics (as shown in Fig.~\ref{fig:Thermodynamics}(B)), reinforcing the link between the weakening spin stiffness and the reorganization of local spin textures.

\begin{figure}[!t]
\centering
\includegraphics[width=8.5cm]{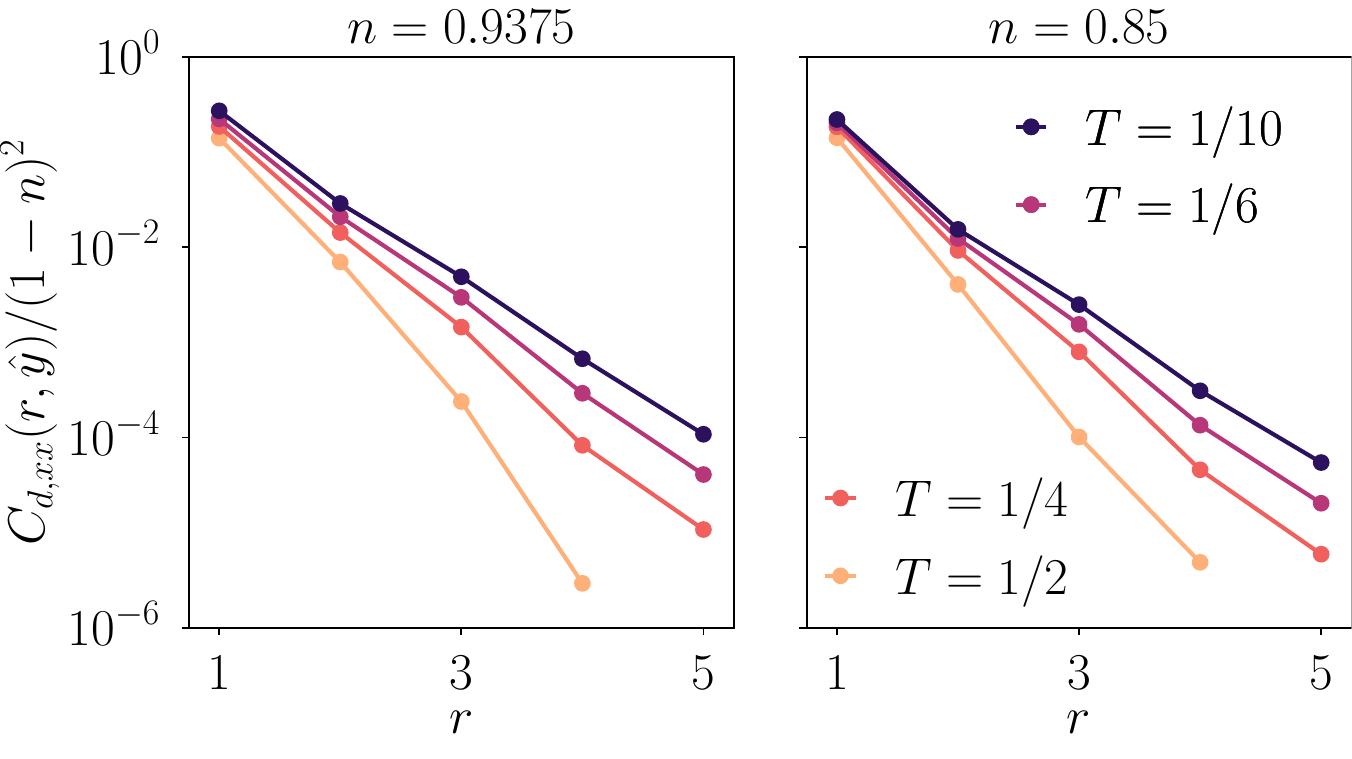}
\caption{\textbf{ $d$-wave correlator -- } $d$-wave correlators at different temperatures being studied decay exponentially.}
\label{fig:dwave}
\end{figure}

Beyond the hole-spin correlators, to probe pairing tendencies we evaluate the $d$-wave correlator. The $d$-wave pairing function is defined as $\Delta_{ij}=(\tilde{c}_{i\downarrow}\tilde{c}_{j\uparrow}-\tilde{c}_{i\uparrow}\tilde{c}_{j\downarrow})/\sqrt{2}$. For simplicity, we focus on the pairing function when site $i$ and site $j$ are the closest neighbor: $\Delta_{\alpha}(\mathbf{r_0})=\Delta_{\mathbf{r}_0,\mathbf{r}_0+\hat{\alpha}}$ in which $\hat \alpha=\hat x,\hat y$. The $d$--wave correlator is then defined as
\begin{equation}
C_{d,\alpha\beta}(r,\hat{\mathbf{n}})\;\equiv\;\Big\langle \Delta^\dagger_{\alpha}\!\big(\mathbf{r}_0+\mathbf{r}\big)\,
\Delta_{\beta}\!\big(\mathbf{r}_0\big)\Big\rangle,
\label{eq:pairing}
\end{equation}
where $\hat{\mathbf{n}}=\mathbf{r}/\|\mathbf{r}\|$. To the lowest order, $C_{d,\alpha\beta}\propto(1-n)^2.$ In Fig.~\ref{fig:dwave}, we show the $d$-wave correlation functions normalized by $(1-n)^2$ at different temperatures and different distances when $\alpha=\beta$.
These data demonstrate that, in the thermodynamic limit and for our parameter set, $J=0.5$ and $n=0.85$ or $0.9375$, $d$-wave pair correlations are strictly short-ranged at accessible temperatures, and there is no sign of any Kosterlitz–Thouless-like growth.

In Appendix \ref{app:additional_correlators}, we show results for two other kinds of correlators: the five-point correlator consisting of four spin
 operators on the nearest-neighbor sites surrounding a central hole projector, and the electron momentum distribution function.

\section{Discussion}\label{sec:conclusion}

In this work we provide thermodynamic-limit, finite-temperature benchmarks for the two-dimensional square-lattice $t$--$J$ model down to $T/t\simeq 0.1$ and up to hole doping $1-n\lesssim 0.25$ using purification-iPEPS. Three main messages emerge. 
First, the bulk thermodynamics show a coherent crossover structure upon cooling: a broad spin-entropy feature in the specific heat $C_v(T)$ associated with the build-up of short-range antiferromagnetism, a pronounced maximum in the uniform susceptibility $\chi(T)$ at a scale $T^\ast$ that shifts and weakens with doping, and a shallow enhancement of the compressibility $\kappa(\mu,T)$ that develops upon cooling in a similar doping window. 
Second, dopant-conditioned correlators reveal the corresponding short-distance microscopic response: single holes behave as magnetic polarons that weaken nearby nearest-neighbor AFM bonds relative to the unconditional background, whereas nearest-neighbor hole pairs generate a cooperative response that reinforces antiferromagnetism on the plaquette edge parallel to the pair, consistent with a local reduction of exchange frustration when two holes move together. Third, over the same parameter window the $d$-wave pairing correlator remains strictly short-ranged, indicating that in the regime accessed here the dominant finite-$T$ reorganization is magnetic and local, preceding any growth of superconducting coherence. These findings are timely in light of recent quantum-gas-microscope studies of the doped Hubbard model, where a susceptibility saturation/maximum versus temperature and a compressibility maximum versus doping develop upon cooling and have been used to map out a pseudogap crossover from thermodynamics~\cite{Chalopin2024,Kendrick2025Pseudogap}. Our results provide complementary thermodynamic-limit benchmarks for the $t$--$J$ model and, crucially, supply a real-space mechanism for how such thermodynamic anomalies can arise from local spin-texture reorganization around dopants without requiring macroscopic phase separation at finite temperature. A key limitation is that our two-site checkerboard iPEPS ansatz enforces translation symmetry up to two sublattices and therefore cannot represent stripe order or other longer-ranged symmetry breaking. Accessing the stripe regime and its interplay with pairing will require larger unit cells. Within the experimentally relevant temperature and doping range explored here, the combined thermodynamic and conditional-correlation data support a simple conclusion: the finite-temperature dopant-induced reorganization of the $t$--$J$ model is predominantly short-ranged and magnetic, while superconducting correlations remain short-ranged.

The data used in the main text of this manuscript can be found in Ref.~\cite{UJ/VTBGIO_2025}. The data used in the Appendix can be given upon request.

\begin{acknowledgments}
We gratefully acknowledge Polish high-performance computing infrastructure PLGrid (HPC Centers: ACK Cyfronet AGH) for providing computer facilities and support within the computational grant no. PLG/2024/017860.

This research was
%
%
funded by the National Science Centre (NCN), Poland, under projects 2024/55/B/ST3/00626 (J.D.) 
2019/35/B/ST3/01028 (Y.Z.), 
and 2020/38/E/ST3/00150 (M.M.R.). 
The research was also supported by a grant from the Priority Research Area DigiWorld under the Strategic Programme Excellence Initiative at Jagiellonian University (MMR,JD).
A.S. acknowledges the Alexander von Humboldt Foundation for support under the Humboldt Research Fellowship.
A.S. thanks Alexander Wietek for extensive discussions and initial help with the METTS implementation, and the entire team of the experimental collaboration of Ref.~\cite{Chalopin2024} for stimulating discussions, particularly Thomas Chalopin.
\end{acknowledgments}


\appendix

\section{Numerical Details}
\label{app:numerical_details}

\begin{figure}[!t]
\centering
\includegraphics[page=5]{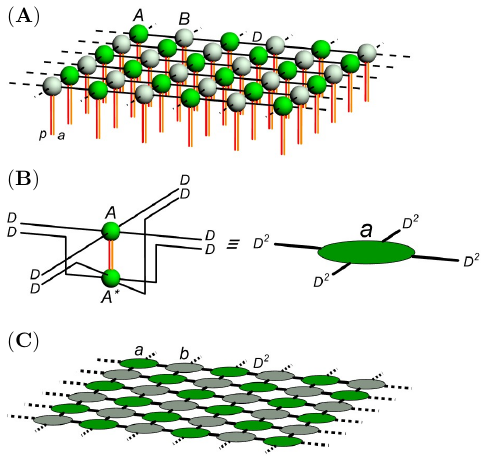}
\caption{\textbf{iPEPS purification -- }
In (A), the checkerboard iPEPS ansatz $\ket{\psi}$ for the thermal purification with two sublattice tensors $A$ and $B$. Here, the red/orange lines are the physical/ancilla indices and the black lines are bond indices of dimension $D$ contracting NN tensors. The dashed lines connect with the rest of the infinite tensor network that is not shown. The bond dimension $D$ limits correlations that can be accommodated by iPEPS and is used as a refinement parameter. All presented results are converged in increasing $D$.
In (B), a contraction of the tensor $A$ with its conjugate $A^*$ to make a double iPEPS tensor $a$.
A similar contraction of $B$ and $B^*$ makes a double $b$.
In (C), the double iPEPS tensors are contracted to make the norm squared $\braket{\psi}{\psi}$. In Fig.~\ref{fig:CTMRG}, the norm is evaluated by the multiplet CTMRG.
Here, any line intersection implies a fermionic swap gate~\cite{CorbozSWAP,10.21468/SciPostPhysLectNotes.25}.
}
\label{fig:iPEPS}
\end{figure}

In this section, we introduce the methods that were used in this manuscript. The thermal Gibbs state $\rho\propto e^{-\beta H}$ was represented by its purification $\ket{\psi}$ as: $\rho={\rm Tr}_a\ket{\psi}\bra{\psi}$. Here $\beta$ is the inverse temperature and the partial trace is over ancillas~\cite{verstraete2004matrix, czarnik12, CzarnikDziarmagaCorboz}. At $\beta=0$ the purification was initialized as a product over lattice sites with the same maximally entangled state between the physical site ($p$) and its ancilla copy ($a$) at every site $j$:
\be
\ket{\psi(0)} \propto \prod_j
\left(
\ket{0_{j,p}}\ket{0_{j,a}}+\ket{\uparrow_{j,p}}\ket{\uparrow_{j,a}}+\ket{\downarrow_{j,p}}\ket{\downarrow_{j,a}}
\right).
\ee
The initial state was evolved as $\ket{\psi(\beta)}\propto e^{-\beta H/2}\ket{\psi(0)}$ with $H$ acting on physical sites.
The purification was represented by the infinite projected entangled pair state (iPEPS) ansatz in Fig.~\ref{fig:iPEPS}(A). The checkerboard ansatz is translationally invariant up to the sublattices $A$ and $B$. We used tensors with $U(1)\times U(1)$ symmetry to make them more compact and reduce computational costs. The evolution in $\beta$ was performed with the second order Suzuki-Trotter decomposition. After every nearest-neighbor Trotter gate, which increases the bond dimension between the NN sites, the affected NN iPEPS tensors were initially truncated using the environment-assisted truncation (EAT) \cite{Hubbard_Sinha} and then further optimized using the neighborhood tensor update (NTU) \cite{ntu,Hubbard_Sinha}. Finally, the expectation values were calculated using a multiplet corner transfer matrix renormalization group (mCTMRG). mCTMRG is the key technical innovation of this work presented below.

\subsection{Neighborhood tensor update (NTU)}

The NTU was introduced in Ref.\cite{ntu}. It was used to optimize the truncated NN tensors after each NN Trotter gate. NTU approximates the environment of the updated tensors using an exact contraction of a cluster of their neighboring tensors. The exactness makes the Gram-Schmidt (G-S) metric tensor, used to minimize the error of the truncation, manifestly Hermitian and non-negative. The NN tensors were initially truncated using the environment assisted truncation (EAT) ~\cite{Hubbard_Sinha} that, unlike SVD, takes into account the non-trivial G-S metric, although the metric is approximated by a Hermitian non-negative product that ignores some of the loopiness of the neighboring tensor cluster. Nevertheless, the initial EAT truncation is more accurate than SVD. This is important because it predefines $U(1)\times U(1)$ sector sizes of the truncated NN tensors that cannot be corrected by the following optimization of the tensors in the full NTU G-S metric. The optimization brings back the loopiness ignored by EAT, but without changing the sector sizes.

The choice of the neighboring cluster affects the efficiency of the computations. On the one hand, the bigger is the cluster, the more accurate is the G-S metric, and the limited iPEPS bond dimension is employed more efficiently. On the other hand, exact contraction of a bigger cluster is more expensive. A systematic study in the appendices of Ref. \cite{King2025} suggests that the optimal cluster size correlates with the range of quantum correlations in the simulated state. In this work, we used the NN+ type of cluster defined there.

\subsection{Multiplet corner transfer matrix renormalization group (mCTMRG)}

The norm squared $\braket{\psi}{\psi}$ is a prerequisite to calculating expectation values. It is a double-layer iPEPS in Fig.~\ref{fig:iPEPS}(C) made of double iPEPS tensors in Fig.~\ref{fig:iPEPS}(B). Its every column/row is a transfer matrix and CTMRG~\cite{Corboz13_su3hc} obtains its leading eigenvectors approximated by matrix product states with an environmental bond dimension $\chi$. In the multiplet CTMRG the approximation is also controlled by a truncation parameter $\epsilon_{C}$ and a multiplet resolution $r_m$.
A set of corner matrices and edge tensors represents the surrounding tensor environment. There are 4 corner tensors $C^{a(b)}_{TL}$, $C^{a(b)}_{TR}$, $C^{a(b)}_{BR}$ and $C^{a(b)}_{BL}$ representing, respectively, the top-left, top-right, bottom-right and bottom-left environment corners, and 4 edge tensors $T^{a(b)}_{L}$, $T^{a(b)}_{T}$, $T^{a(b)}_{R}$ and $T^{a(b)}_{B}$ representing, respectively, the left, top, right and bottom environment edges. The corner and edge tensors can be initialized with their environmental bond dimension $1$, and then they are repeatedly updated until convergence. Each update at first enlarges their environmental dimension and then truncates it back.

\begin{figure}[!t]
    \includegraphics[page=4]{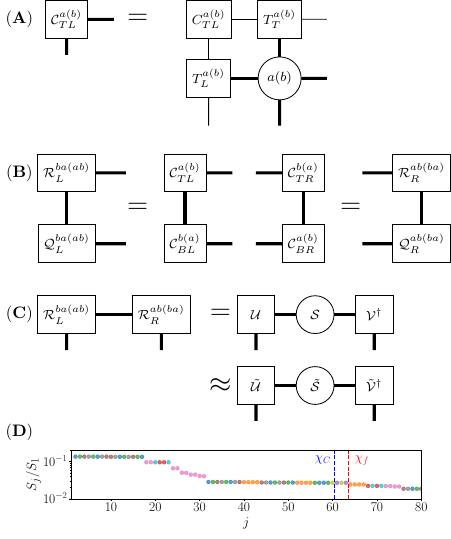}
    \caption{{\bf Multiplet CTMRG -- }
    (A)
    Enlarged corner tensor $\mathcal{C}_{TL}^{a(b)}$. Other enlarged corner tensors are defined in the similar fashion. Here $a(b)$ are the double iPEPS tensors defined in Fig.~\ref{fig:iPEPS}(B), but no colors are used in the planar diagrams here.
    (B)
    QR decomposition. Here $\mathcal{Q}_{L(R)}^{ba(ab)}$ is a unitary matrix.
    (C)
    Singular value decomposition and truncation of the singular values $S$. The truncated matrices are the ones with a tilde.
    (D)
    A typical example of the singular value spectrum $S$. Different colors represent singular values from different blocks. The spectrum has degenerate multiplets. The blue dashed line indicates the position of the cut at $\chi_C$ while the red dashed line indicates the position of the cut at $\chi_f$ after considering the multiplet structure.
    }\label{fig:CTMRG}
\end{figure}

Each update at first defines enlarged corners as in Fig.~\ref{fig:CTMRG}(A). To update CTM horizontally, we first perform QR decompositions as in Fig.~\ref{fig:CTMRG}(B), where $Q^{ab(ba)}_{R}$ and $Q^{ba(ab)}_{L}$ are unitary matrices and $R^{ba(ab)}_{L}$ and $R^{ab(ba)}_{R}$ are upper triangle matrices. Then, we do singular value decomposition (SVD) as in Fig.~\ref{fig:CTMRG}(C).
The diagonal $\mathcal{S}$ has $\chi_0$ singular values sorted in the non-increasing order: $\mathcal{S}_1\geq \mathcal{S}_2\geq \cdots\geq \mathcal{S}_{\chi_0}$. Instead of simply truncating $\mathcal{S}$ to a predefined number of singular values, we noticed that $\mathcal{S}$ has a multiplet structure that needs to be dealt with carefully when it comes to the truncation. To do this, we initially truncate $\mathcal{S}$ to $\chi_i$ such that $\chi_i$ is the largest number satisfying $\mathcal{S}_{\chi_i}/\mathcal{S}_1\geq\epsilon_C$ and $\chi_i\leq\chi$. Next, we find the largest $\chi_f$ such that $\chi_f\geq\chi_i$ and $1-\mathcal{S}_{\chi_f}/\mathcal{S}_{\chi_i}<r_m$. By including a complete multiplet during truncation, we improve the stability of CTMRG. Finally, the CTM tensors can be updated horizontally using the protocol described in Ref.~\cite{10.21468/SciPostPhysLectNotes.25}. Updating CTM vertically is done in a similar fashion.

To monitor if the mCMTRG is converged, we calculate the differences of singular values of the 8 corner tensors $C^{X}_{Y}$ in which $X\in\{A,B\}$ and $Y\in\{TL,TR,BL,BR\}$ of the $i^{\mathrm{th}}$ iteration. Note that since we are using symmetric tensors throughout the calculation, the corner tensors are block-wise. We denote the singular values of the corner tensor $C_{Y}^{X}$ of the $i^{\mathrm{th}}$ and  $(i+1)^{\mathrm{th}}$iteration as $s_{jk}^{X,Y,i}$ in which $j$ enumerates different blocks and $k$ enumerates the singular values in the $j^\mathrm{th}$ block. The singular values of each block are sorted in non-increasing order. The singular values of each iteration are normalized such that $\max_{jk}s_{jk}^{X,Y,i}=1$. The error between $i^{\mathrm{th}}$ and ${(i+1)}^{\mathrm{th}}$ iteration is defined by
\begin{equation}
    \epsilon_{C,i}=\max_{XY}\big(\sum_j\|s^{X,Y,i}-s^{X,Y,i+1}\|_2\big).
\end{equation}
The CTMRG is then terminated when $\epsilon_{C,i}\leq\epsilon_C$ or $i>i_{\mathrm{max}}$. Since we do not need to evaluate any expectation values at each iteration of the mCTMRG to check the convergence, mCTMRG improves both the stability and efficiency of CTMRG.
In this paper, we set $\epsilon_C=10^{-5}$. The resolution of multiplets $r_m$ is set as the $i^{\mathrm{th}}$ gap $\delta_i=\mathcal{S}_i-\mathcal{S}_{i+1}$ in which $i$ is the smallest index that satisfies $i\geq\chi$ and $\delta_i\geq \mathcal{S}_{i+1}$.

We used $\chi=80,100,120$ for $D=25$ and checked the energy difference between the last two iterations. The energies converge at the level of $10^{-6}\sim10^{-8}$ even at low temperatures. After considering the multiplet structure, the final bond dimension $\chi_f$ can be $50\%$ larger than $\chi$.

\section{Chemical Potential}\label{app:chemical_potential}

\begin{figure}[!b]
\centering
\includegraphics[width=8.5cm]{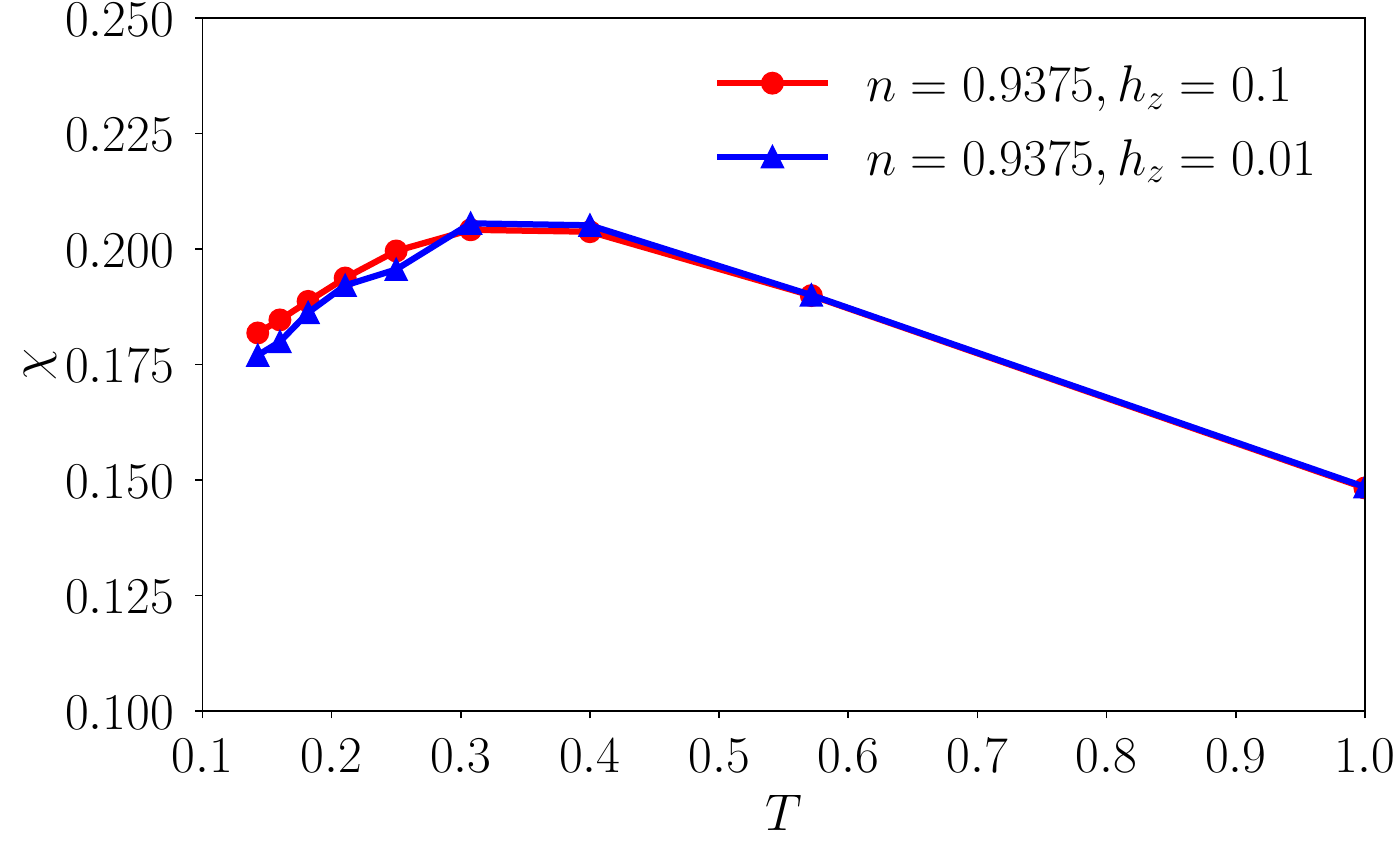}
\caption{\textbf{Comparison of magnetic susceptibility between different biased fields -- } Red lines and the blue line shows the results of biased field $h_z=0.1$  and $h_z$=0.01 at doping $n=0.9375$ respectively.}
\label{fig:appendix_sus}
\end{figure}

For a fixed density $n$, the chemical potential is plotted versus temperature $T=1/\beta$ in Fig.~\ref{fig:T_vs_mu}. Similar to Refs.~\cite{jaklivc2000,rigol2007}, we find that the chemical potential has a linear dependence on temperature for high temperature and high filling,
\be
\mu(T) \approx \mu(T=0) + \alpha T.
\ee
Every filling seems to be associated with a characteristic temperature where they start deviating from the high entropy linear behavior that may be connected with pseudogap phase or significant hole clustering. For high doping i.e., for filling of around $n\sim 0.8$, the behavior starts looking more quadratic
\bea
\mu(T) \approx \mu(T) + \beta T^2
\eea
which signals the emergence of coherent quasiparticle excitations reminiscent of Fermi liquid behavior.


\section{Comparison between the magnetic susceptibility of different biased fields}\label{app:different_field}
\label{sec:comparison_field}

As mentioned in Sec. \ref{sec:thermo}, we used two different biased fields $h_z$ to calculate the magnetic susceptibility. In Fig. \ref{fig:appendix_sus}, we show the magnetic susceptibility using $h_z=0.1$ and $h_z=0.01$. The results agree with each other quantitatively. Note that the weaker field may be deeper in the linear regime at the cost of being more vulnerable to numerical noise. 


\section{Comparison between the magnetic susceptibility and specific heat of iPEPS and METTS}\label{app:ipeps_vs_metts}
\label{sec:comparison}

\begin{figure}[!b]
\centering
\includegraphics[width=8.5cm]{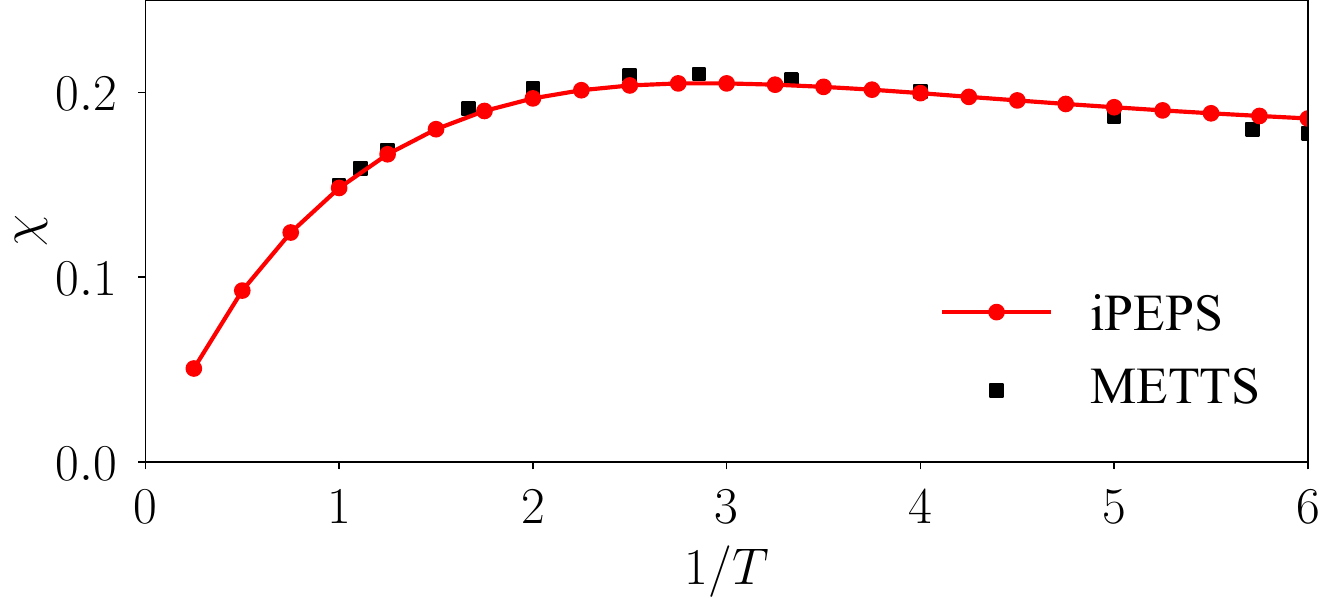}
\includegraphics[width=8.5cm]{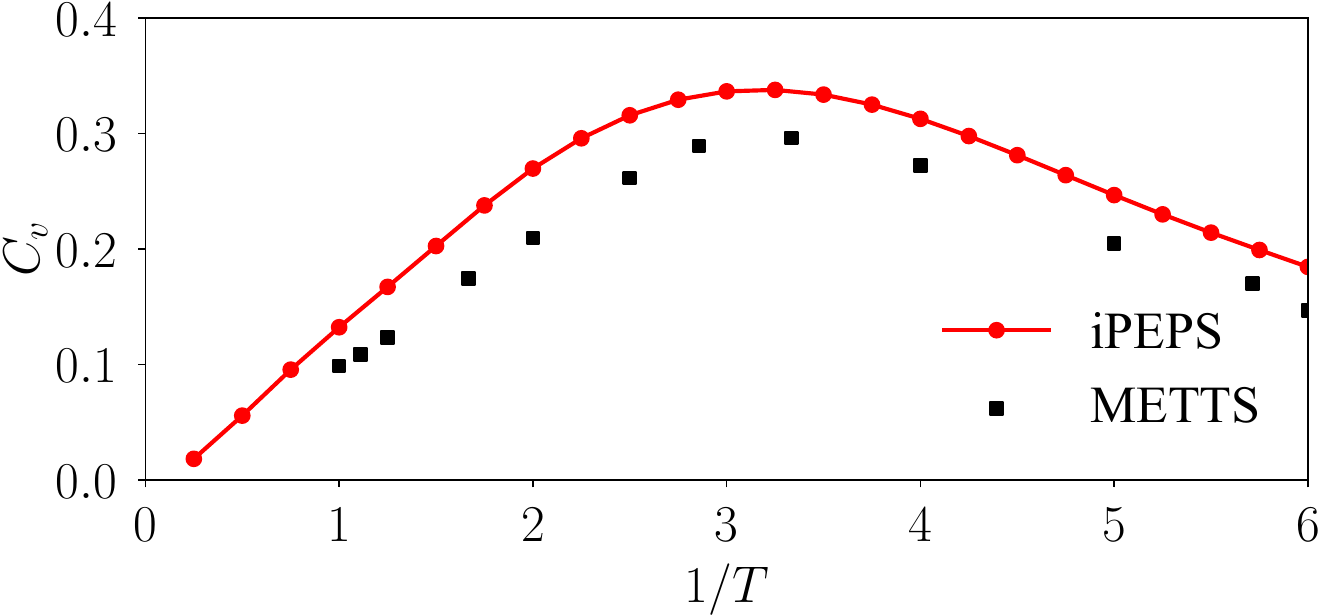}
\caption{\textbf{Comparison between iPEPS and METTS -- } Red lines are the results of iPEPS and the black squares are the results of METTS done of acylinder of length $24$ and width $4$. The upper panel shows the results of magnetic susceptibility, while the lower one shows the results of specific heat. Here, we chose filling $n=0.9375$.}
\label{fig:iPEPSvsMETTS}
\end{figure}
\begin{figure*}[t]
\centering
\includegraphics[width=0.85\textwidth]{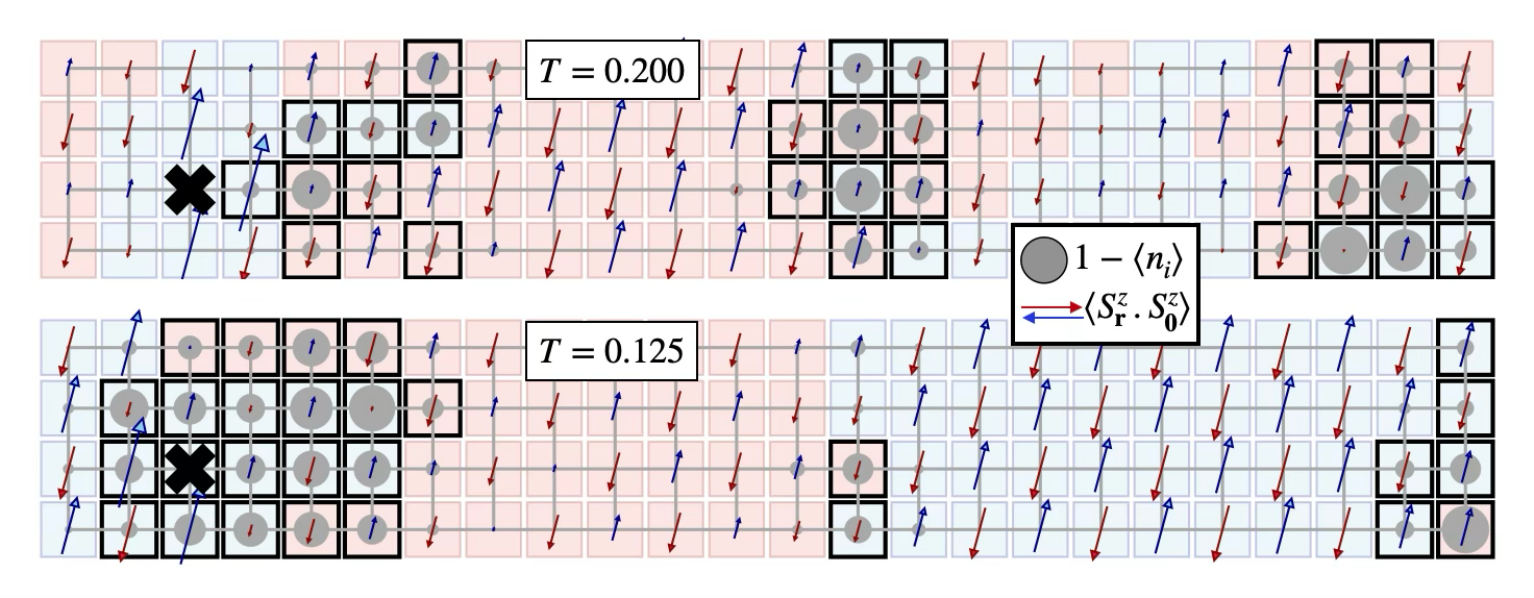}
\caption{\textbf{METTS snapshots for the $t$-$J$ cylinder -- }
Here $n=0.9375$ at $T=0.200$ (top) $T=0.125$ (bottom). Lattice sites are numbered by coordinate $r=(x,y)$ on a $24\times4$ cylinder. Grey circles encode the local hole density $n_h(r)=1-\langle n_r\rangle$; the circle radius is proportional to $n_h(r)$ with a fixed reference scale across panels. Arrows depict the equal-time spin correlator $\langle S^z_{0}\,S^z_{r}\rangle$ with respect to a fixed reference site $r=0$ marked by a black cross; arrow length is proportional to $|\langle S^z_{0}\,S^z_{r}\rangle|$ and color encodes the sign (blue for $+$, red for $-$). Squares are color-filled by the sign of the staggered correlator $(-1)^{x+y}\langle S^z_{0}\,S^z_{r}\rangle$ to delineate N\'eel  domains; adjacent domains differ by a $\pi$ phase shift in sign of antiferromagnetic correlations and thus appear in alternating colors. Sites outlined in black satisfy the hole-rich criterion of Eq.~\eqref{eq_app_threshold}.
}
\label{app_snapshots}
\end{figure*}
In this section, we choose two thermodynamic quantities, the magnetic susceptibility and specific heat, to compare iPEPS with another popular finite temperature method, i.e., minimally entangled typical thermal states (METTS) applied to matrix product states. The MPS-METTS calculations used \texttt{METTS.jl} (\url{https://github.com/awietek/METTS.jl}), built on top of the ITensor library~\cite{10.21468/SciPostPhysCodeb.4-r0.3,10.21468/SciPostPhysCodeb.4}. For METTS simulation, we used the same parameters as in the main text on a $24 \times 4$ lattice (open along length $L=24$, periodic along width $W=4$). For iPEPS simulation, we set the bond dimension of iPEPS tensor to be $D=25$ and the bond dimension of the CTMRG tensors to be $\chi=100$. We compare the results at filling $n=0.9375$ in Fig.~\ref{fig:iPEPSvsMETTS}. As for magnetic susceptibility, the METTS result agrees with iPEPS result. The specific heat follows the same trend but is larger in iPEPS, primarily because the thermodynamic-limit, boundary-free iPEPS state supports more long-wavelength energy fluctuations per site than the finite $24\times4$ cylinder, raising $C_{v}(T)$ via the fluctuation–dissipation relation.

\section{Charge clustering at intermediate temperatures in the $t$--$J$ model via METTS}
\label{app:clustering}
\begin{figure}[t]
\centering
\includegraphics[width=0.95\columnwidth]{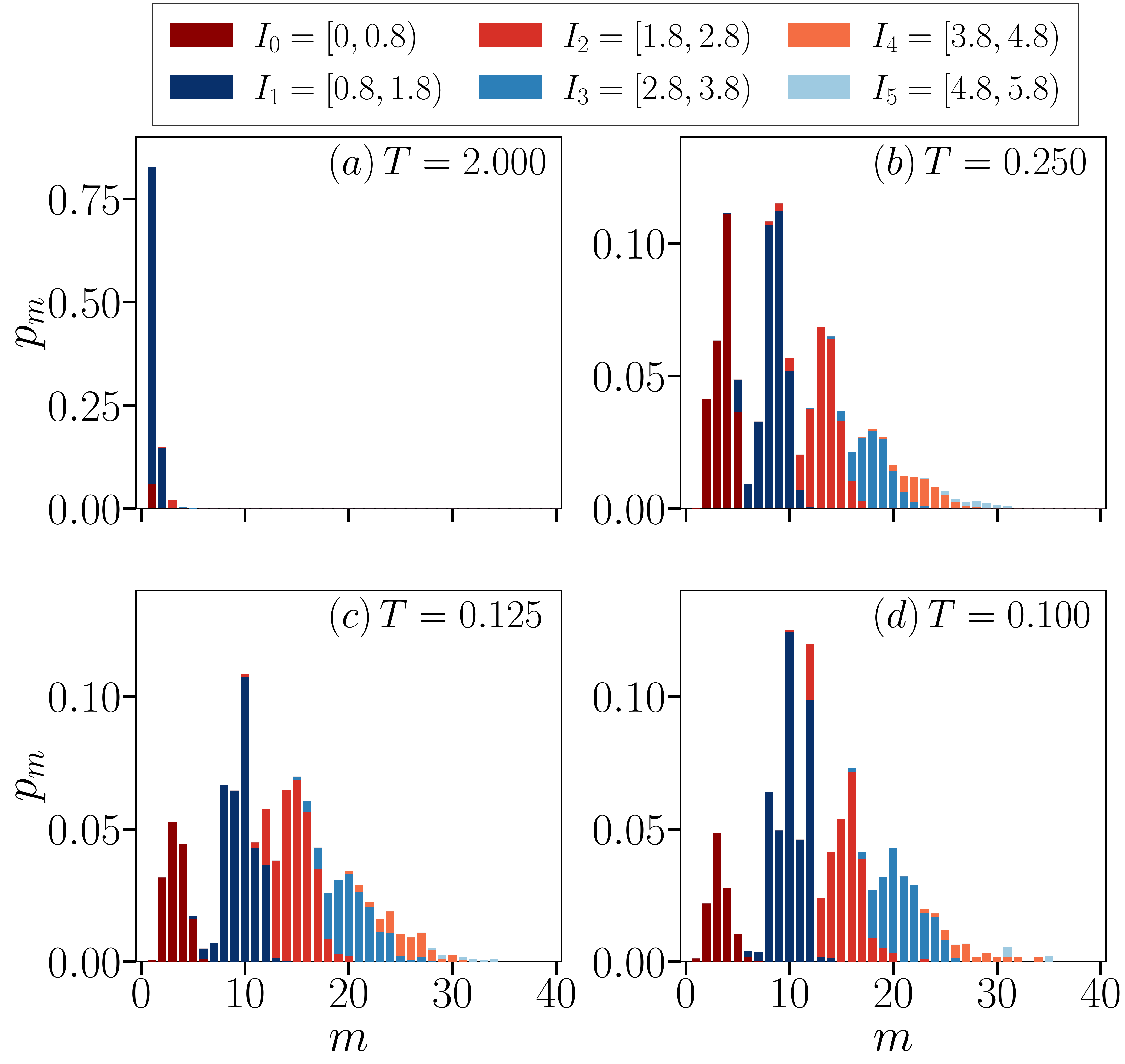}
\caption{\textbf{Cluster–size histograms from METTS snapshots resolved by hole mass -- }
Stacked bars of the density-weighted probabilities $p_m$ versus cluster size $m$ for the $24\times4$ $t$-$J$ cylinder at $T\in\{2.000,\,0.250,\,0.125,\,0.100\}$.
Colors encode the cluster hole mass $\rho(\mathcal C)=\sum_{r\in\mathcal C}n_h(r)$ binned into the non-overlapping windows
$\mathcal{I} = [0,0.8)$, $[0.8,1.8)$, $[1.8,2.8)$, $[2.8,3.8)$, $[3.8,4.8)$, $[4.8,5.8)$; the total height at each $m$ equals $p_m$. Nearest-neighbor connectivity is used with periodic wrap across the width (cylinder) and open boundaries along the length.}
\label{fig_cluster_histogram}
\end{figure}

Here, we investigate whether fluctuating charge clusters are the microscopic origin of the enhanced charge compressibility $\kappa=\partial n/\partial\mu$ reported in the main text for $T<0.25$, following a similar claim for the Hubbard model~\cite{sinha2024forestalled}. We have already found in Fig.~\ref{fig:iPEPSvsMETTS} that METTS and iPEPS agree on critical thermodynamic quantities like specific heat and magnetic susceptibility down to temperatures $T=1/6$. The local hole density is defined as $n_h(r)=1-\langle n_r\rangle$. We gather thermal states as sampled by minimally entangled typical thermal states (METTS)~\cite{stoudenmire10, wietek2021} on a $24\times4$ cylinder (open along $L = 24$, periodic along $W = 4$) at $J=0.5$, filling $n=0.9375$, and maximum bond dimension $D_{\max}=2000$. For each snapshot we define the standard deviation over the lattice $\Lambda$ (set of all sites) as $\sigma_{n_{h}}=\sqrt{\frac{1}{|\Lambda|}\sum_{r\in\Lambda}\big(n_h(r)-(1-n)\big)^2}$, and label a site hole–rich if it exceeds the adaptive threshold
\begin{equation}
n_h(r)>n_h^{\mathrm{th}}\equiv (1-n)+c\sigma_h,\qquad c=0.3.
\label{eq_app_threshold}
\end{equation}
In Fig.~\ref{app_snapshots}, we have two representative METTS snapshots. The radius of the grey circle at site $r$ is proportional to $n_h(r)$ with a fixed reference radius across panels. The arrow at $r$ has length proportional to $|\langle S^z_{0}S^z_{r}\rangle|$ and color set by its sign, where $r=0$ is the marked reference site. The square color shows the sign of $(-1)^{x+y}\langle S^z_{0}S^z_{r}\rangle$, so Néel domains appear as contiguous regions of the same color. Black outlines indicates hole rich sites.

Let $\mathcal E=\{\,r\in\Lambda\mid n_h(r)>n_h^{\mathrm{th}}\}$ be the set of hole rich sites. We partition $\mathcal E$ into disjoint clusters $\mathcal C$ defined as nearest–neighbor connected components on the cylinder,
\begin{equation}
\mathcal E=\dot{\bigcup}_{\mathcal C}\mathcal C.
\end{equation}
Each cluster carries a size $m=|\mathcal C|$ and a hole mass
$\rho(\mathcal C)=\sum_{r\in\mathcal C}n_h(r)$. Aggregating clusters over snapshots at fixed $T$, we form the density-weighted cluster size distribution
\begin{equation}
p_m=\frac{\displaystyle\sum_{\mathcal C:\,|\mathcal C|=m}\rho(\mathcal C)}{\displaystyle\sum_{\mathcal C}\rho(\mathcal C)}\,,\qquad \sum_m p_m=1.
\end{equation}
To identify which amounts of charge supply the weight at each $m$, we resolve
$p_m$ by coarse–graining the cluster mass $\rho(\mathcal C)$ into a set of
non–overlapping, integer–anchored windows,
$\mathcal I=\{I_0,I_1,I_2,\ldots\}$,
\begin{equation}
I_0=[0,0.8),\ \
I_k=[\,k-0.2,\ k+0.8\,),\ \ (k\ge1)
\label{windows}
\end{equation}
and decompose
\begin{equation}
p_m=\sum_{I\in\mathcal I}p_m^{(I)},\qquad
p_m^{(I)}=\frac{\displaystyle\sum_{\mathcal C:\,|\mathcal C|=m,\ \rho(\mathcal C)\in I}\rho(\mathcal C)}{\displaystyle\sum_{\mathcal C}\rho(\mathcal C)}.
\end{equation}
Fig.~\ref{fig_cluster_histogram} presents the stacked histograms of $p_m$ versus $m$ at $T=2.000,\,0.250,\,0.125,\,0.100$, where the color of each bar segment encodes the window $I\in\mathcal I$ and the total bar height equals $p_m$. These windows are asymmetric intervals of width $1.0$ chosen so that the
observed lobe maxima of $p_m$ vs. $m$ fall roughly near the middle of the
corresponding window that labels their dominant total hole mass. Importantly, this coloring does not modify $p_m$ itself. At $T=2.000$ the distribution is nearly exponential and dominated by $m=1$ with $\rho(\mathcal C)\in[0.8,1.8)$. Upon cooling, the weight shifts to larger $m$ and broad lobes emerge. The mass-resolved stacks show that the successive lobes of $p_{m}$ draw most of their weight from successive integer-anchored windows $I_{k}$, i.e., the clusters tend to accumulate approximately one additional hole as $m$ increases from lobe to lobe. In other words it is the cluster mass that is near-integer on a per lobe basis, not an enforced binning artifact. This pattern is compatible with an AFM background that penalizes extended domain walls, so aggregating carries step wise (one by one) lowers the exchange cost. Note that this lobe-like feature exists mainly at low dopings when the background antiferromagnetism is strong. The evolution of $p_m$ mirror the idea of forestalled phase separation crossover seen in the Hubbard model~\cite{sinha2024forestalled}: clusters proliferate and grow but remain finite at intermediate temperature regimes above the onset of stripes.

\section{Additional correlators}
\label{app:additional_correlators}
\begin{figure}[!t]
\centering
\includegraphics[width=7.5cm]{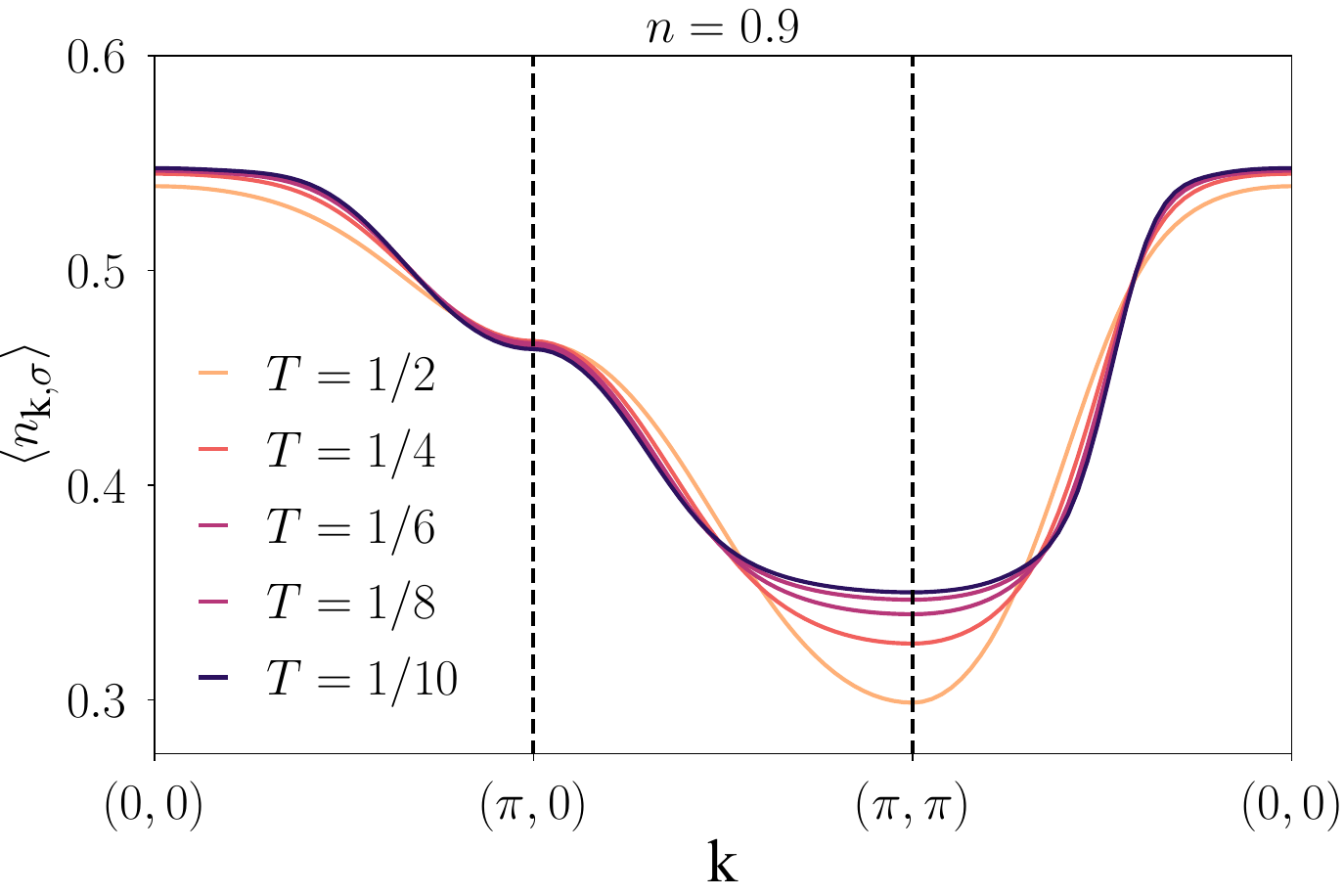}
\caption{\textbf{Momentum distribution -- } $n_{\mathbf{k}\sigma}$ along the high-symmetry path $\Gamma(0,0)\to X(\pi,0)\to M(\pi,\pi)\to\Gamma$ for the $t$--$J$ model at filling $n=0.9$ (with $J=0.5$), shown for several temperatures $T$.}
\label{fig:k-space}
\end{figure}

We also studied some fermionic correlators on our iPEPS tensor network. Unlike the bosonic correlators, we need to take special care of the bond crossing when doing the fermionic one to follow the anti-commutation relation of the fermionic operators. In the no double occupancy subspace we use  the Fourier modes of projected fermionic operators, $\tilde c_{\mathbf{k}\sigma}=\frac{1}{\sqrt{N}}\sum_{j}e^{-i\mathbf{k}\cdot\mathbf{r}_j}\,\tilde c_{j\sigma}$ and momentum distribution function \cite{Singh1992},
\begin{equation}
 n_{\mathbf{k}\sigma}\equiv\big\langle \tilde c_{\mathbf{k}\sigma}^\dagger \tilde c_{\mathbf{k}\sigma}\big\rangle =\frac{1}{N}\sum_{i,j}e^{i\mathbf{k}\cdot(\mathbf{r}_i-\mathbf{r}_j)}\big\langle \tilde c_{i\sigma}^\dagger \tilde c_{j\sigma}\big\rangle
\end{equation}
Fig.~\ref{fig:k-space} shows the momentum distribution $n_{k\sigma}$ along the high symmetry path $(0,0) \rightarrow (\pi,0) \rightarrow (\pi, \pi) \rightarrow (0,0)$ for $T=1/2, 1/4, 1/6, 1/8, 1/10$. A clear minimum occurs at $(\pi, \pi)$; its value increases as $T$ is lowered, i.e. the $(\pi,\pi)$ dip becomes shallower on cooling. The values near $(0,0)$ vary only weakly with $T$, and the curves are essentially on top of each other at $(0,\pi)$. No sharp Fermi-step discontinuity is observed at these finite temperatures; the profile remains smooth with a Fermi-surface-like gradient. Short-range AFM correlations with $\mathbf{Q} = (\pi, \pi)$ hybridize $k$ and $k+\mathbf{Q}$, which reduces but does not eliminate the occupation difference; correspondingly $n(\pi, \pi)$ increases on cooling while remaining below $n(0,0)$.

\begin{figure}[!t]
\centering
\includegraphics[width=7.5cm]{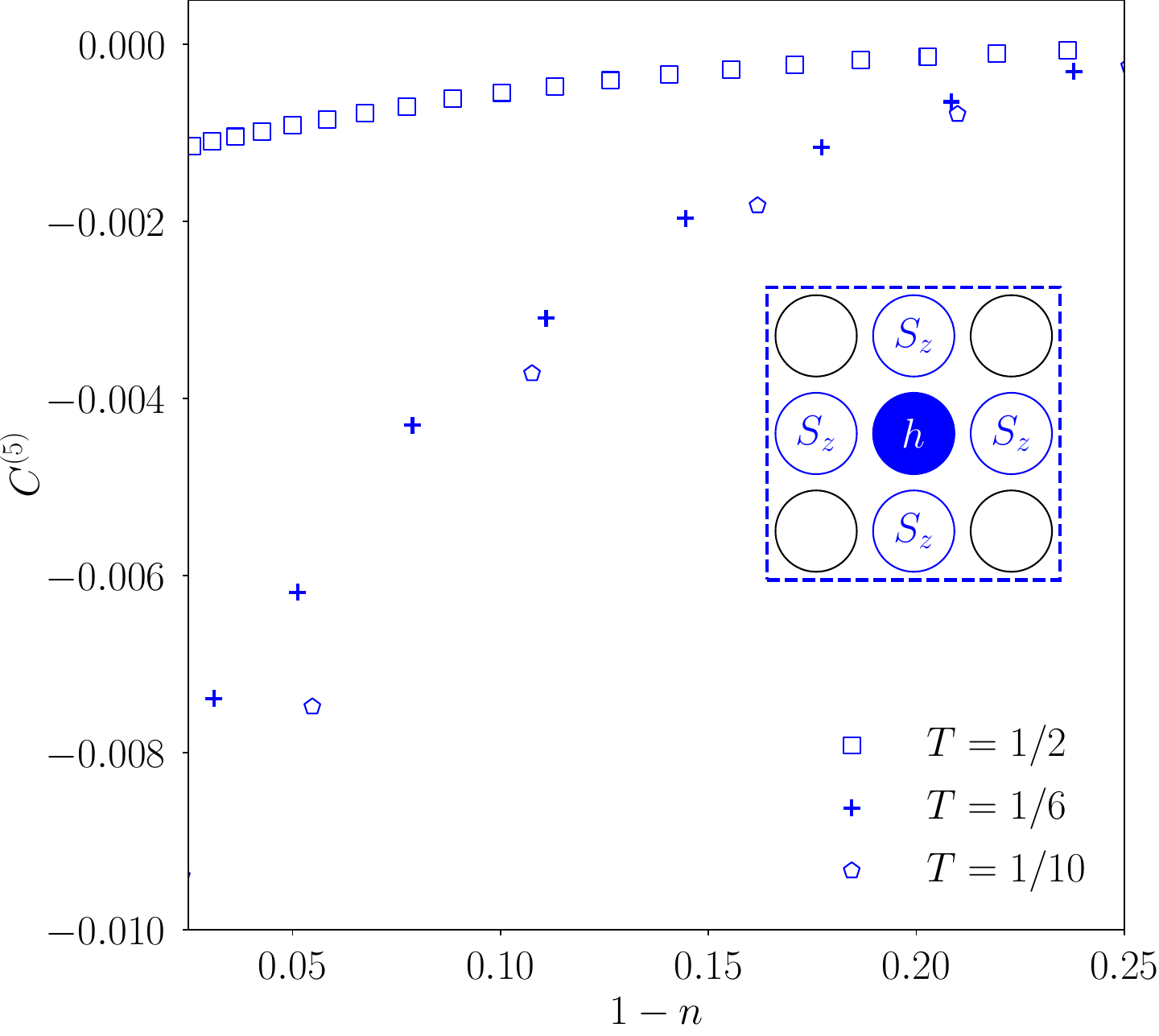}
\caption{\textbf{5P correlator.} Five-point correlator $C^{(5)}$ for the geometry shown in the inset for different temperatures.}
\label{fig:5P}
\end{figure}

Here, we demonstrate an example of calculating a 5P correlator. We first generalize the ${B_{ab}}$ introduced in Eq.~\eqref{eq:C3_def} from the main text to $\mathcal{B}_{abcd}\equiv S_a^zS_b^zS_c^zS_d^z$. Then, following a similar logic as in Eq.~\eqref{eq:C3_def} from the main text, we define a 5P correlator as shown in the inset of Fig.~\ref{fig:5P}. The spin operators are sitting at the 4 nearest-neighbor sites of the hole operator. For simplicity, we denote this particular kind of conditional correlator as $C^{(5)}$. Fig.~\ref{fig:5P}, shows $C^{(5)}$ at $T=1/2,1/6,1/10$.

\bibliography{ref.bib}

\end{document}